\documentstyle[aps]{revtex}

\begin{document}
\title{A speculative relation between the cosmological constant and 
the Planck mass}
\author{S. Hsu$^{*}$ \& A. Zee$^{**}$}
\address{
$~~$\\
$^{*}$Institute of Theoretical Science\\
University of Oregon\\
Eugene, OR 97403\\
hsu@duende.uoregon.edu \\
$~~$ \\
$^{**}$Kavli Institute for Theoretical Physics\\
University of California\\
Santa Barbara, CA 93106\\
zee@kitp.ucsb.edu\\
$~~$\\
}
\maketitle

\begin{abstract}
We propose the relation $M_{\Lambda }\sim (M_{Pl}M_{U})^{\frac{1}{2}}$ where
$M_{\Lambda }$, $M_{Pl},$ and $M_{U}$ denote the mass scale associated with
the cosmological constant, the gravitational interaction, and the size of
the universe respectively.
\end{abstract}


\bigskip

The cosmological constant problem has been exacerbated by the discovery of
dark energy. In this note we offer a speculative discussion of the magnitude
of the cosmological constant. Since quantum gravity has yet to be
formulated, we cannot give a complete step-by-step derivation of our result,
which should be regarded as merely a conjecture or a suggestion for a
possible avenue of attack on this enormous problem. However, in our opinion
the proposal is sufficiently intriguing that it merits a brief note.

Below the quantum gravity scale, the action for the world $S(g,\Phi )$ (here
$\Phi $ denotes all matter fields generically) starts with the purely
gravitational terms $S(g,\Phi )=\int d^{4}x\sqrt{-g}(-\Lambda
+M_{Pl}^{2}R+\cdots ).$ We use a convention in which the cosmological
constant $\Lambda $ has the dimension of an energy density and the metric
has the $(+,-,-,-)$ signature. In this convention, the observed cosmological
constant is positive. The cosmological ``constant'' $\Lambda $ is to be
thought of as the sum of a ``bare'' cosmological term $\Lambda _{0}$ and
contributions due to matter fields, so that $\Lambda =\Lambda
_{0}+V(<\varphi >,$ $...)$ where $V$ includes quantum contributions to the
leading term, e.g., as computed by Coleman and Weinberg. The generic bosonic
field $\varphi$ may of course be a composite of fermionic fields $\psi$. The
important point for us here is merely that $\Lambda $ is not a fixed
parameter from the fundamental action, but a variable which depends on
various vacuum expectation values and whose value we can obtain by
extremization.

One aspect of the cosmological constant puzzle is that $-\Lambda $ couples
to the spacetime volume $\int d^{4}x\sqrt{-g}$ of the universe: the
cosmological constant is a sort of Lagrange multiplier for the spacetime
volume of the universe. At present, the only reasonable interpretation of
the ``spacetime volume of the universe'' appears to be $L^{4},$ where $L$
denotes some length scale characteristic of the universe, say the Hubble
radius $10^{10}$ years $=$ 3 $\times $ $10^{17}$ seconds (see below for more
discussion). If so, then we expect that after we integrate out the
gravitational and matter fields in
\begin{equation}
Z=\int Dg\,D\Phi ~e^{iS(g,\Phi )}  \label{functional}
\end{equation}
we obtain an effective action $S_{eff}$ with the leading term $-\Lambda
L^{4}.$ This term goes smoothly through 0 as $\Lambda $ goes through 0, from
positive values to negative values.

We need a term which does not behave smoothly as $\Lambda $ goes through 0.
What we would like to suggest here is that quantum gravity produces a term
of the form $-M_{Pl}^{4}/\Lambda ,$ so that the effective action becomes
\begin{equation}
S_{eff}\sim -(\Lambda L^{4}+M_{Pl}^{4}/\Lambda )+\{{\rm \Lambda ~indep.~terms%
}\}  \label{effaction}
\end{equation}
Extremizing this action gives us
\begin{equation}
\Lambda \sim (M_{Pl}/L)^{2}  \label{res}
\end{equation}

We find it convenient to define a mass scale $M_{\Lambda }$ associated with
the cosmological constant by $\Lambda =M_{\Lambda }^{4}$ and a mass scale $%
M_{U}$ associated with the universe by $M_{U}=L^{-1},$ sort of a Compton
mass of the universe. Thus, the mass scale responsible for the cosmological
constant is given rather pleasingly by the geometric mean of the Planck mass
$M_{Pl}$ and the ``cosmic mass'' $M_{U}$%
\begin{equation}
M_{\Lambda }\sim (M_{Pl}M_{U})^{\frac{1}{2}}  \label{geome}
\end{equation}

Numerically, $M_{U}\sim (3\times 10^{17}{\rm \sec })^{-1}\sim 2\times
10^{-33} {\rm eV}$. With $M_{Pl}\sim 10^{19}$ GeV we obtain $M_{\Lambda
}\sim 4.6\times 10^{-3}$ eV, which happens to be in order of magnitude
agreement with the fourth root of the observed dark energy density.

The difficulty is of course to justify the term $-M_{Pl}^{4}/\Lambda .$
Minic and Horava\cite{wolfpara} have proposed an entropic factor $e^{{\cal S}%
}$ of the form ${\cal S}\sim M_{Pl}^{4}/\Lambda $ to account for the
enormous density of states when the cosmological constant is given by $%
\Lambda .$ Here, however, we are doing a Minkowskian path integral and it is
not clear how the entropic factor is analytically continued to $%
e^{-iM_{Pl}^{4}/\Lambda }.$

Some years ago, Baum\cite{baum} and Hawking\cite{hawk} studied Euclidean
quantum gravity and concluded that the cosmological constant $\Lambda $ is
governed by a probability distribution of the form $e^{M_{Pl}^{4}/\Lambda }$
peaking at $\Lambda =0$ for $\Lambda >0.$ Again, we need to analytically
continue if we want to invoke this result. Readers familiar with the
Baum-Hawking analysis might also object that we may be double counting here
since the Euclidean version of the first term in (\ref{effaction}) appears
in an intermediate step there. In any case, it is far from clear whether the
Baum-Hawking analysis itself is sensible, e.g., due to issues concerning
Euclidean quantum gravity.

Perhaps we could simply regard (\ref{effaction}) as an effective
Landau-Ginzburg type action for $\Lambda ,$ not excluding terms with
negative powers of $\Lambda $ since there aren't any good reasons to exclude
them. This action might result from integrating out all degrees of freedom
with masses greater than $M_U$, leading to an effective theory containing
only the size of the universe (the long-distance component of the metric)
and the cosmological constant as degrees of freedom. (We ignore the photon
and other massless particles - it is not clear in any case whether they can
have correlation lengths of order $L$.) For some reason, the effective
action is dominated by the two terms shown in (\ref{effaction}).

The result for the value of the cosmological constant in (\ref{geome})
reminds us of the seesaw mechanism for neutrino masses, which was
conjectured without a firm theoretical derivation, with the neutrino masses
determined by the clash between two disparate mass scales. Here also, $%
M_{\Lambda }$ is supposed to result from the clash between the largest mass
scale $M_{Pl}$ and the smallest mass scale $M_{U}$ known in fundamental
physics. Amusingly, $M_{\Lambda }$ may be of the same order of magnitude as
the neutrino mass\cite{hz1}. (Is this just a coincidence or the result of
something more profound?)

The value for $\Lambda $ obtained in (\ref{res}) happens to be the same as
the upper bound on $\Lambda $ obtained by Cohen, Kaplan, and Nelson\cite{ckn}
and by others\cite{thomas}. Their heuristic analysis invokes ideas from
holography, but one simple interpretation is that the value of $\Lambda $ is
such that the universe is on the verge of becoming a black hole. Recall that
an object of mass $M$ and size $R$ becomes a black hole unless $GM<R$, that
is $M<M_{Pl}(M_{Pl}R).$ Applying this to the entire universe (whether or not
this makes sense remains to be seen) with $M=\Lambda L^{3}$ and $R=L$ we
obtain $\Lambda <(M_{Pl}/L)^{2},$ which is to be compared with (\ref{res}).

It is perhaps not surprising that the same relation keeps popping up: if we
believe that $\Lambda $ has something to do with the Planck scale (the
ultraviolet cutoff) and with the size of the universe (the infrared cutoff),
and if one were to try to obtain a value for $\Lambda $ by equating simple
algebraic expressions, one could hardly avoid getting relations of the type
proposed here and in \cite{ckn,thomas} (see also \cite{padm}). Indeed, if we
believe that $M_{U}$ is the only other relevant mass scale besides the
Planck mass then $M_{\Lambda }=M_{Pl}~f(M_{U}/M_{Pl})$ by dimensional
analysis, so the simplest guess would be that $M_{\Lambda }$ is reduced from
$M_{Pl}$ by some power of the suppression ratio $M_{U}/M_{Pl}.$ Here we
suggest that the power is $\frac{1}{2}.$

An ambiguity in this proposal, which also applies to related
holographic proposals, is that if we equate $L$ with the Hubble radius $%
\Lambda $ would depend on time. Hsu\cite{hsu} was the first to point out
that this leads to an unacceptable time dependence. Later, Li\cite{li} made
an intriguing proposal that $L$ is to be equated instead with the size of
the future event horizon:
\begin{equation}
L(t)~\sim ~a(t)\int_{t}^{\infty }\frac{dt^{\prime }}{a(t^{\prime })}~.
\end{equation}
As usual, $a(t)$ denotes the scale factor in the
Robertson-Walker metric. This $L$, which describes the size of the largest
portion of the universe that any observer (even an infinitely long-lived
one) will see, seems a reasonable choice for the ultimate infrared cutoff.
Interestingly, the observational consequences discussed in \cite{li} are
consistent with the current supernova data and an equation of state $w\sim
-1 $. Given our present lack of understanding, we prefer to regard the
precise characterization of $L$ (or $M_{U}$) as a higher order difficulty to
be addressed after a more satisfactory derivation of (\ref{res}).

\bigskip

{\bf {\large Acknowledgments\medskip }}

This work was supported in part by the National Science Foundation under
grant number PHY 99-07949 and by the U.S. Department of Energy under grant
number DE-FG06-85ER40224.\medskip

\newpage

{\bf {\large References\medskip }}

\end{document}